\documentclass[twocolumn,prc,aps,showpacs]{revtex4}
\usepackage{graphicx}

 \def\gsim{\mathrel{\rlap{\lower4pt\hbox{\hskip1pt$\sim$}}
 \raise1pt\hbox{$>$}}}

 \newcommand\la{\langle}
 \newcommand\ra{\rangle}
 \newcommand\beq{\begin{equation}}
 
 \newcommand\eeq{\end{equation}}
 \newcommand\beqn{\begin{eqnarray}}
 \newcommand\eeqn{\end{eqnarray}}
\def\mb{\,\mbox{mb}}
\def\fm{\,\mbox{fm}}
\def\GeV{\,\mbox{GeV}}
\def\TeV{\,\mbox{TeV}}
\def\lsim{\mathrel{\rlap{\lower4pt\hbox{\hskip1pt$\sim$}}
    \raise1pt\hbox{$<$}}}         
\def\gsim{\mathrel{\rlap{\lower4pt\hbox{\hskip1pt$\sim$}}
    \raise1pt\hbox{$>$}}}         
\def\J{J/\Psi}

\def\mb{\,\mbox{mb}}
\def\fm{\,\mbox{fm}}
\def\GeV{\,\mbox{GeV}}
\def\MeV{\,\mbox{MeV}}

\def\s0{\sigma_0(s)}

\def\beq{\begin{equation}}
\def\eeq{\end{equation}}

\def\beqy{\begin{eqnarray}}
\def\eeqy{\end{eqnarray}}

\newcommand{\ber}{\begin{displaymath}}
\newcommand{\eer}{\end{displaymath}}
\newcommand{\bey}{\begin{eqnarray}}
\newcommand{\eey}{\end{eqnarray}}

\def\beq{\begin{equation}}
\def\eeq{\end{equation}}

\def\beqy{\begin{eqnarray}}
\def\eeqy{\end{eqnarray}}

\begin{document}

\title{\bf  Heavy quarkonium production:\\
Nontrivial transition from pA to AA collisions}

\vspace{1cm}

\author{B. Z. Kopeliovich$^{1,2}$}
\author{I. K. Potashnikova$^{1}$}
\author{H. J.~Pirner$^2$}
\author{Iv\'an Schmidt$^1$}
\affiliation{$^1$Departamento de F\'{\i}sica
Universidad T\'ecnica Federico Santa Mar\'{\i}a; and
\\
Instituto de Estudios Avanzados en Ciencias e Ingenier\'{\i}a; and\\
Centro Cient\'ifico-Tecnol\'ogico de Valpara\'iso;\\
Casilla 110-V, Valpara\'iso, Chile\\
$^2$Institut f\"ur Theoretische Physik der Universit\"at,\\
Philosophenweg 19, 69120 Heidelberg, Germany}
\begin{abstract}
\noindent Two novel QCD effects, double color filtering and mutual boosting of the saturation scales in colliding nuclei, affect the transparency of the nuclei for quark dipoles in comparison with proton-nucleus collisions. The former effect increases the survival probability of the dipoles, since color filtering in one nucleus makes the other one more transparent.
The second effect acts in the opposite direction and  is stronger, it makes the colliding nuclei more opaque than in the case of $pA$ collisions. As a result of parton saturation in nuclei the effective scale is shifted upwards, what leads to an increase of the gluon density at small $x$. This in turn leads to a stronger transverse momentum broadening in $AA$ compared with $pA$ collisions, i.e. to an additional growth of the saturation momentum. Such a mutual boosting leads to a system of reciprocity equations, which result in a saturation scale, a few times higher in $AA$ than in $pA$ collisions at the energies of LHC. Since the
dipole cross section is proportional to the saturation momentum squared, the nuclei become much more opaque for dipoles in $AA$ than in $pA$ collisions. For the same reason gluon shadowing turns out to be boosted to a larger magnitude compared with the product of the gluon shadowing factors in each of the colliding nuclei.
All these effects make it more difficult to establish a baseline for anomalous $\J$ suppression in heavy ion collisions at high energies.
\end{abstract}



\maketitle

\section{Introduction}

Nuclear suppression of heavy quarkonia is usually considered as a sensitive hard probe for the properties of the short-living medium produced in heavy ion collisions \cite{satz,kps-psi,k-bnl}.
The main challenge is to discriminate between initial state interactions (ISI), usually identified as cold nuclear matter effects, and final state interaction (FSI), which is related to attenuation of the produced quarkonium in the dense matter created in the nuclear collision. While the latter is the main goal of
the study, experimental information  fully depends on how well we understand the ISI contribution.

Naturally, the ISI dynamics should be studied in proton-nucleus collision, where no dense matter is expected to be produced. The next step, the extrapolation of the results to nuclear collisions, is not that easy. Usually it is done in an oversimplified manner, assuming that a $\bar cc$ pair is produced momentarily inside the nucleus and then attenuates with an unknown absorption cross section on the way out of the nucleus. The $\bar cc$ break-up cross section is fitted to $pA$ data and used to predict
the ISI effects in $AA$ collisions. As is shown below, the break-up cross section is not constant, but steeply rises with the $\bar cc$ energy, besides, it is well known from HERA data.
The most appealing oversimplification, especially at the high energies of RHIC and LHC, is the instantaneous production of $\bar cc$. In reality the production time ranges from tens to thousands fermi at these energies. 
The long production time leads to shadowing which is a high twist effect suppressed by the quark mass squared. Nevertheless, we find that this shadowing effect is stronger than the leading twist gluon shadowing, which produces a rather mild suppression even at LHC in the central rapidity region.

Even if the shadowing and break-up effects in $pA$ collisions are under control , still one cannot predict the cold nuclear effects in $AA$ collisions in a model independent way. The two major phenomena considered in this paper make this impossible. The first one is double color filtering of dipoles propagating simultaneously through two nuclei. As is demonstrated below, the survival probability of a dipole in $AA$ collisions is higher than the product of that in each of the colliding nuclei. 

Another effect changing the properties of the cold nuclear medium in $AA$ collisions, is the boosting of the saturation scales in the nuclei due to mutual multiple interactions in the nuclei. The effect is especially strong at the energies of LHC. The nuclear medium becomes several times more opaque for dipoles compared with its transparency in $pA$ collisions. These two effects essentially modify the 
effects of the ISI stage of heavy ion collisions, which is usually considered as the baseline for search for "anomalous" nuclear suppression of heavy quarkonia.

All calculations throughout this paper are performed in so called "frozen" approximation,
neglecting the size fluctuations of dipoles during their propagation through nuclei.
This approximation is accurate provided that the coherence length for heavy quark production substantially exceeds the size of the nuclei. This condition is satisfied at the energies of RHIC and above.

\section{Proton-nucleus collisions}

First of all one should make sure that the nuclear effects for heavy quarkonium production in $pA$ collisions are understood, as well as the absolute magnitude of the cross section. The cross sections of $\chi$ and $J/\Psi$ production in $pp$ collisions was successfully reproduced in the color singlet model in refs. \cite{kth-psi} and \cite{bl}
respectively.
What has been missed in most of current analyses of $\J$ production in $pA$ collisions, are the coherence effects related to the long time scale of charm production, associated with propagation of a color-octet $\bar cc$ pair through the nucleus prior  the production of a colorless dipole \cite{kth-psi,klt}. This stage results in the heavy quark shadowing, which is of the same order as the effect of break-up of the colorless dipole. Although both are the high twist effects, quantitatively they are more important for currently available data, than the suppression caused by the leading twist gluon shadowing. The latter is frequently miscalculated basing on some of nuclear gluon PDFs,
which rely on either ad hoc or incorrect assumptions (see discussion in \cite{k-bnl}).

\subsection{High twist heavy quark shadowing}

Although the proper time of charm production is short, $t_c^*\sim 1/(2m_c)$, this time rises with $\J$ energy linearly, in the rest frame of the nucleus, 
\beq
t_c\sim\frac{2E}{M_{\J}^2}.
\label{120}
\eeq
Thus, if the energy of the produced $\J$ is sufficiently high, $E\gsim 25(\GeV)\times L(\fm)$, effects of coherence become significant. At the energy of RHIC, $\sqrt{s}=200\GeV$, and positive rapidities, $t_c>12\fm$. For $\Upsilon$ production $t_c$ at RHIC is rather short, but becomes much longer than the nuclear size at the energies of LHC. In what follows we assume the coherence time to be much longer than the nuclear size, unless otherwise specified.

The nuclear  suppression caused by coherence can be interpreted as  high twist shadowing in the process of $\bar cc$ pair production by a projectile gluon. The $\bar cc$ is produced coherently in multiple interactions of the projectile gluon with target nucleons. If the $\J$ energy is sufficiently high,
so that $t_c\gg R_A$, one can neglect the dipole size fluctuations during propagation through the nucleus, i.e. treat the dipoles being "frozen" by Lorentz time dilation.

Since the production amplitude is convoluted with the charmonium wave function, one can assume with good accuracy an equal sharing of the total longitudinal momentum between $c$ and $\bar c$. 
In what follows we rely on the saturated parametrization of the dipole cross section \cite{gbw}
\beq
\sigma_{\bar cc}(r_T,x_2)=\sigma_0\left[1-e^{-r_T^2/r_0^2(x_2)}\right],
\label{130}
\eeq
where $\sigma_0=23.03\mb$; $r_0(x_2)=0.4\fm\times(x_2/x_0)^{0.144}$; $x_0=3.04\times10^{-4}$;
$x_2=e^{-y}\,\sqrt{\la M_{\bar cc}^2\ra+\la p_T^2 \ra}\biggl/\sqrt{s}$. The $\bar cc$ invariant mass distribution predicted by the color singlet model leads to $\la M_{\bar cc}^2\ra=2M_{\J}^2$ \cite{kps-psi}. The measured $\la p_T^2\ra=4\GeV^2$.

The amplitude of $\bar cc$ production at a point with impact parameter $b$ and longitudinal coordinate $z$ inside the nucleus, averaged over the dipole size, reads \cite{kth-psi},
\beqn
&&S_{pA}(b,z)=\int d^2r_T\,W_{\bar cc}(r_T)
\label{140}
\\ &\times&
\exp\left[-{1\over2}\sigma_{\bar ccg}(r_T)T_-(b,z)
-{1\over2}\sigma_{\bar cc}(r_T)T_+(b,z)\right].
\nonumber
\eeqn
Here $T_-(b,z)=\int_{-\infty}^z dz'\rho_A(b,z')$;~ $T_+(b,z)=T_A(b)-T_-(b,z)$, and $T_A(b)=T_-(b,\infty)$. 
According to \cite{kth-psi} shadowing for $\bar cc$ production over the nuclear thickness $T_-(b,z)$ occurs with the shadowing cross section
corresponding to a 3-body dipole, gluon and $\bar cc$, which for equal momenta of $c$ and $\bar c$ equals to $\sigma_{\bar ccg}(r_T)={9\over4}\sigma_{\bar cc}(r_T/2)-{1\over8}\sigma_{\bar cc}(r_T)$.

The survival probability amplitude Eq.~(\ref{140}) should be squared and integrated over the coordinate of the production point. Then the nuclear ratio reads,
\beq
R_{pA}={1\over A}\int d^2b\int\limits_{-\infty}^{\infty}dz\,
\left|S_{pA}(b,z)\right|^2.
\label{145}
\eeq

We rely on the form of nuclear effects for production of the $P$-wave charmonium $\chi_2$ derived in \cite{kth-psi}. The mechanism of $\J$ production is more complicated, but the general structure of shadowing corrections should be similar, because the overlap of the initial narrow size distribution of the perturbatively produced $\bar cc$ pair with the large size wave function of the final charmonium is dominated by the small initial size $\sim 1/m_c$, whether the charmonium is $\chi$ or $\J$.
In addition, $\chi$ decays contribute about $30\%$ to $\J$ production. 

In the regime of long $t_c\gg R_A$ the weight factor in (\ref{140}) has the form \cite{kth-psi},
\beq
W_{\bar cc}(r_T)\propto K_0(m_c r_T)\,r_T^2\,\Psi_{\J}(r_T),
\label{180}
\eeq
where one factor $r_T$ comes from the amplitude of $\bar cc$ production, and another one either from the amplitude of gluon radiation in the case $\J$ production, or from the radial wave function of $\chi_2$.
Further, we assume the wave function to have the oscillatory form, $\Psi_{\J}(r_T)\propto\exp[-r_T^2/2\la r_{\J}^2\ra]$ with $\la r_{\J}^2\ra=2/m_c\omega$, $m_c=1.5\GeV$, and $\omega=300\MeV$ \cite{kz91}.

\begin{figure}[htb]
\begin{center}
\includegraphics[width=6cm]{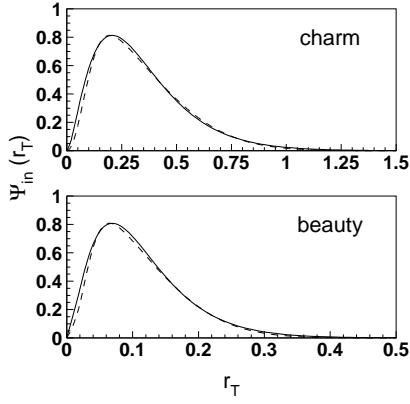}
\end{center}
\caption{\label{psi-rT} Comparison of the left-hand (solid curves) and right-hand (dashed) sides of Eq.~(\ref{182}).
The top and bottom panels are for charm and beauty respectively.}
 \end{figure}

To simplify the calculations we use for $\Psi^c_{in}(r_T)=r_T^2\,K_0(r_Tm_c)$ the parametrization proposed in \cite{kz91}, 
\beq
\Psi^c_{in}(r_T)\approx 
const\times\left[e^{-r_T^2/a_c^2}-e^{-r_T^2/b_c^2}\right],
\label{182}
\eeq
where $a_c=0.496\fm$, $b_c=0.11\fm$. The comparison of the two functions Eq.~(\ref{182}) plotted in Fig.~\ref{psi-rT} demonstrates that the approximation is rather accurate within the range of $r_T$ we are interested in. 

Thus, we present the weight factor $W_{\bar cc}$ Eq.~(\ref{180}) in the form,
\beq
W_{\bar cc}(r_T)=\frac{1}{\pi(r_1^2-r_2^2)}\left[e^{-r_T^2/r_1^2}-e^{-r_T^2/r_2^2}\right],
\label{185}
\eeq
where $r_1^2=a_c^2/(1+a_c^2/2\la r_{\J}^2\ra)$, and $r_2^2=b_c^2/(1+b_c^2/2\la r_{\J}^2\ra)$.

In small-$r_T$ approximation the dipole-nucleon cross sections in (\ref{140}) have the simple form,
$\sigma_{\bar cc}(r_T,x_2)=C(x_2)\,r_T^2$, and $\sigma_{\bar ccg}(r_T,x_2)={7\over16}C(x_2)\,r_T^2$.
So, the integration can be performed analytically, and we arrive at
\beq
S_{pA}(b)=\frac{1}{r_1^2-r_2^2}\left[r_1^2\,S_{pA}^{(1)}(b)- r_2^2\,S_{pA}^{(2)}(b)\right]
\label{195}
\eeq
where for $i=1,2$
\beq
S_{pA}^{(i)}(b)=
\left[1+{1\over2}C\,r_i^2\left({7\over16}T_-(b,z)
+T_+(b,z)\right)\right]^{-1}
\label{197}
\eeq
Notice that due to color transparency the nuclear medium is more transparent than in the Glauber model. Moreover, the amplitude Eq.~(\ref{197}) does not decrease with nuclear thickness exponentially, but as a power.

Finally, we integrate the attenuation factor Eq.~(\ref{195}) squared over the coordinates of the production point 
and arrive at the nuclear ratio, which has the form,
\beqn
R_{pA}&=&\left(\frac{r_1^2}{(r_1^2-r_2^2}\right)^2R^{(1)}_{pA}+
\left(\frac{r_2^2}{(r_1^2-r_2^2}\right)^2R^{(2)}_{pA}
\nonumber\\ &-&
2\left(\frac{r_1 r_2}{(r_1^2-r_2^2}\right)^2R^{(12)}_{pA},
\label{205}
\eeqn
where for $i=1,2$
\beqn
R^{(i)}_{pA} &=& {1\over A}
\int d^2b\,T_A(b)
\left[1+{r_i^2\over2}\,C\,T_A(b)\right]^{-1}
\nonumber\\ &\times&
\left[1+{7r_i^2\over32}\,C\,T_A(b)\right]^{-1}\!\!\!\!,
\label{150}
\eeqn
and
\beqn
&&R^{(12)}_{pA}={1\over A}\,\frac{32}{9C(r_1^2-r_2^2)}
\int d^2b
\label{155}\\ &\times&
\ln\left(\frac{\left[1+{r_1^2\over2}C\,T_A(b)\right]
\left[1+{7r_2^2\over32}C\,T_A(b)\right]}
{\left[1+{r_2^2\over2}C\,T_A(b)\right]
\left[1+{7r_1^2\over32}C\,T_A(b)\right]}\right).
\nonumber
\eeqn

\begin{figure}[htb]
\begin{center}
\includegraphics[width=7cm]{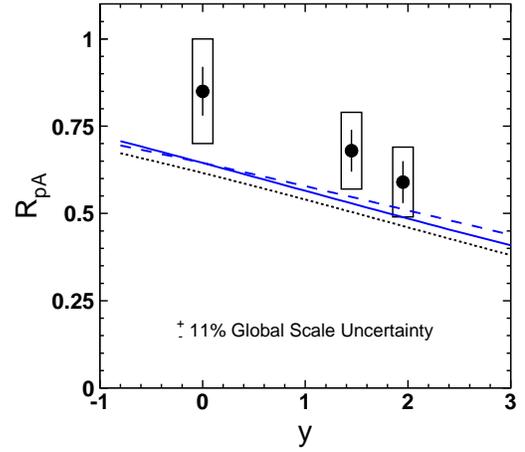}
\end{center}
\caption{\label{d-A-y} (Color online) Dotted curve presents nuclear suppression of $\J$ as function of rapidity in $pA$ collisions calculated with the analytic formula Eq.~(\ref{155}). Dashed curve presents more accurate numerical calculations  with the full dipole cross section Eq.~(\ref{130}).
Solid curve is corrected for gluon shadowing. Data 
are for $d$-$Au$ collisions at $\sqrt{s}=200\GeV$ \cite{phenix-dA}.}
 \end{figure}

With this equations we calculated the nuclear ratio $R_{A/p}(y)$, Eq.~(\ref{205}). We rely on the realistic Woods-Saxon parametrization for the nuclear density \cite{ws}.  
The results at $\sqrt{s}=200\GeV$ are depicted as function of rapidity in Fig.~\ref{d-A-y} by dotted curve.
We see that the steep rise of the break-up cross section $\sigma_{\bar cc}(r_T,E_{\bar cc})$ with energy (it triples from $y=0$ to $y=2$)  explains well the observed rapidity dependence of nuclear suppression. The calculations should not be continued far to negative rapidities, since the regime of long coherence length
breaks down there. Besides, additional mechanisms, which cause a nuclear enhancement at negative rapidities, must be added. 

We also tested how accurate is the result of analytic integration Eq.~(\ref{205}) and performed numerical calculation using the full equation (\ref{140}) with the dipole cross section Eq.~(\ref{130}). The results is plotted by dashed curve in Fig.~\ref{d-A-y}. It is pretty close to the previously calculated dotted curve. Both agree with data within large errors and normalization uncertainty.

\begin{figure}[htb]
\begin{center}
\includegraphics[width=7cm]{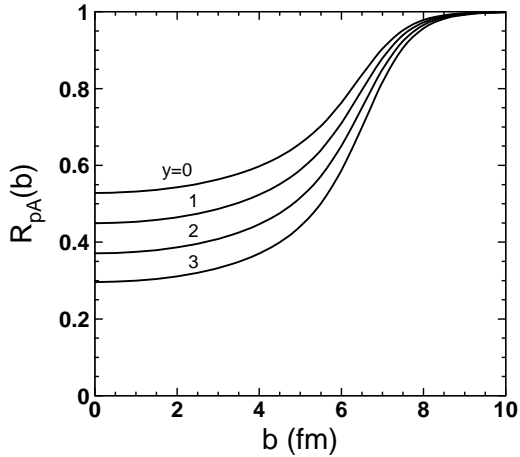}
\end{center}
\caption{\label{d-A-b} $b$-dependence of the nuclear ratios for $\J$ produced with rapidities $y=0,\ 1,\ 2,\ 3$ in $pAu$ collisions at $\sqrt{s}=200\GeV$.}
 \end{figure}

In Fig.~\ref{d-A-b} we present the impact parameter dependence of nuclear suppression for $\J$ produced with different rapidities in $p$-$Au$ collisions at RHIC.
As expected, the strongest dependence on rapidity comes from most central collisions.

\begin{figure}[htb]
\begin{center}
\includegraphics[width=7cm]{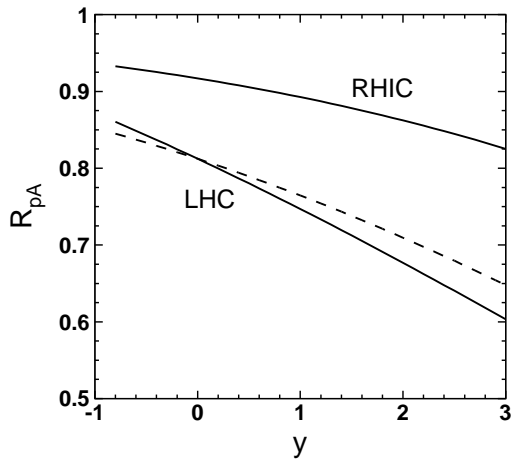}
\end{center}
\caption{\label{upsilon} Nuclear suppression of $\Upsilon$ production as function of rapidity in $pA$ collisions at $\sqrt{s}=200\GeV$ (upper curve, gold) and $\sqrt{s}=5.5\TeV$ (two bottom curves, lead). 
The bottom dashed curve includes only the effects of $\bar bb$ dipole break-up and high twist shadowing of beauty production, solid curve is corrected for gluon shadowing.}
 \end{figure}

We also performed calculations for nuclear effects in $\Upsilon$ production. The only difference with charmonium production is the heavier quark mass, $m_b=4.5\GeV$ and new values of parameters in the parametrization Eq.~(\ref{182}), $a_b=0.162$, $b_b=0.037$. How accurate is this parametrization for bottom quarks is demonstrated in the bottom panel of Fig.~\ref{psi-rT}.
The results for nuclear suppression of $\Upsilon$ produced 
on lead at the energies of RHIC,
$\sqrt{s}=200\GeV$, and LHC,  $\sqrt{s}=5.5\TeV$, as function of rapidity are depicted in Fig.~\ref{upsilon}. 
The upper solid and bottom dashed curves are calculated at $\sqrt{s}=200\GeV$ and $5.5\TeV$ respectively. Notice that in the former case the "frozen" approximation is not well justified at $y=0$, where $t_c\sim 4\fm$. 

\subsection{Leading twist gluon shadowing}

In terms of the Fock decomposition gluon shadowing for $\bar cc$ production is associated with higher Fock states $|\bar ccg\ra$, etc.
First, one should evaluate the kinematic condition for gluon shadowing, $t_c^{\bar ccg}\gsim R_A$, where $t_c^{\bar ccg}$ is the coherence time, or the lifetime of a $\bar ccg$ fluctuation in a gluon.
This time can be related to the Ioffe time as,
\beq
t_c^{\bar ccg}=\frac{P_g}{xm_N},
\label{220}
\eeq
where the factor $P_g\approx0.1$ was evaluated in \cite{krt2} and found to be scale-independent.
Its smallness is caused by the large intrinsic transverse momenta of gluons in hadrons, supported by numerous evidences in data \cite{kst2,spots}.
Thus, shadowing for gluons onsets at  $x\lsim 0.01$, which is a small $x$ value than for quarks.
As the result, no gluon shadowing is possible for charmonium production in any of fixed target experiments performed so far \cite{brahms,k-bnl}.

Even at the energy $\sqrt{s}=200\GeV$ the values of $x_2$ defined in Eq.~(\ref{130}) are too large for gluon shadowing, $0.024>x_2>0.0033$ within the measured rapidity interval $0<y<2$.  We rely upon the NLO analysis  \cite{DS} of DIS data, which suggests a very weak gluon shadowing, as is depicted in Fig.~(6) of \cite{DS}. Such a weak shadowing is in good agreement with the theoretical predictions \cite{kst2}. The nuclear ratio presented by dashed curve  in Fig.~\ref{d-A-y} corrected for gluons shadowing at $Q^2=10\GeV^2$ \cite{DS} (see Fig.~6 in \cite{DS}), is depicted by solid curve. We see that the effect of gluon shadowing is indeed vanishingly small. Even at the energy of LHC, $\sqrt{s}=5.5\TeV$ and $y=0$ gluon shadowing according to \cite{DS,kst2} is extremely small, only $3\%$  at $x_2=5.5\times10^{-3}$, and will be neglected in what follows.

Apparently, for $\Upsilon$ production gluon shadowing is weaker than for charmonia. No gluon shadowing affects $\Upsilon$ production at the energies of RHIC, as is depicted by the upper curve in Fig.~\ref{upsilon}, since the values of $x_2$ are too large. The effect of  gluon shadowing at the energy of LHC is is visible and the full calculation is presented by the bottom solid curve. It was calculated using the gluon nPDF of \cite{DS} at  $Q^2=100\GeV^2$.

\section{Transition from $pA$ to $AA$}\label{transition}

At fist glance one might extrapolate the nuclear effects from $pA$ to $AA$ collisions in a  straightforward way: $R_{AA}(\vec b,\vec \tau)=R_{pA}(\vec\tau)\times R_{pA}(\vec b-\vec\tau)$, where $\vec b$ is the impact parameter of nuclear collisions,  and $\J$ is produced at impact parameter $\vec\tau$. Indeed, such a "data driven" procedure was used in \cite{phenix-dA,cassagnac} to predict the cold nuclear matter effects in nuclear collisions based on the measurements of $b$-dependence of nuclear suppression in $pA$. There are several reason, however, which make such a transition model dependent.

\subsection{Double-color-filtering}

As was discussed above, the survival probability of a $\bar cc$ dipole propagating through a nucleus is subject to color transparency. This effect in $AA$ collisions turns out to be non-factorizable.
This can be illustrated on the following example. Let a $\bar cc$ dipole of transverse separation $r_T$ propagate through a slice of nuclear medium of thickness $T_A$
with survival probability $S_{pA}(r_T)=\exp(-C\,r_T^2\,T_A)$. To clarify the appearance of the effect we rely on the simplified size distribution function $W(r_T)\propto\exp[-r_T^2/\la r_T^2\ra]$ adjusted to the distribution Eq.~(\ref{180}) with $\la r_T^2\ra=0.045\fm^2$.
Then the nuclear attenuation factor takes the form,
\beq
S_{pA}= \int d^2r_T\,W(r_T)\,S_{pA}(r_T)=
\frac{1}{1+C\la r_T^2\ra\,T_A}.
\label{240}
\eeq

Naively,  one could expect $S_{AA}(b)=
S_{pA}^2$. Such a factorization is valid only in the dipole representation for a given dipole size,
$S_{AA}(r_T)=S_{pA}^2(r_T)=
\exp(-2\times Cr_T^2 T_A)$. Factor 2 is introduced because the dipole is simultaneously attenuated by both nuclei.
Now we can repeat the above averaging over dipole size and compare the result (left) with the conventional recipe (right),
\beq
S_{AA}=\frac{1}{1+2\,C\la r_T^2\ra\,T_A}\ 
\Leftrightarrow\  \frac{1}{\bigl[1+C\la r_T^2\ra\,T_A\bigr]^2}
\label{260}
\eeq
We see that the two absorption factors are quite different, especially for $C\la r_T^2\ra T_A\gsim1$.
The source of the difference is color filtering. Namely, the mean transverse size of a $\bar cc$ wave packet propagating through a nucleus is getting smaller, since large-size dipoles are filtered out (absorbed) with a larger probability \cite{zkl,bbgg}. Such a dipole with a reduced mean size  penetrates more easily through
the second colliding nucleus, compared with a $pA$ collision. The mutual color filtering makes both nuclei more transparent.

\begin{figure}[htb]
\begin{center}
\includegraphics[width=6.5cm]{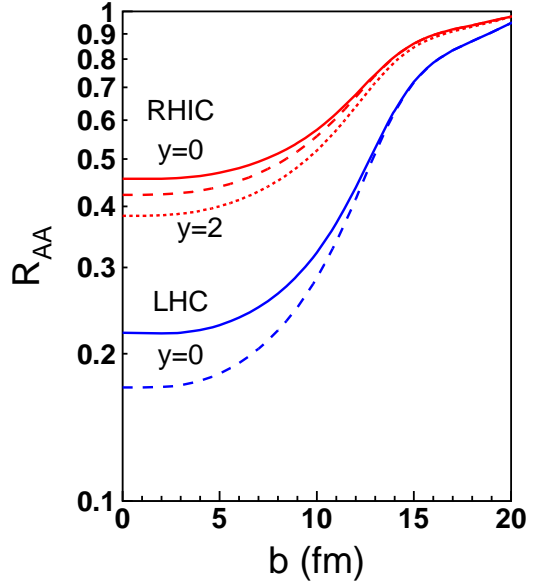}
\end{center}
\caption{\label{A-A-b} (Color online) Effects of double-color-filtering. $\J$ suppression by ISI effects in $Au$-$Au$ collisions at $\sqrt{s}=200\GeV$ as function of $b$. The upper and bottom pairs of curves (solid and dashed) correspond to $y=0$ and energies $\sqrt{s}=200\GeV$ and $5.5\TeV$ respectively. 
Solid and dashed curves present the results at $y=0$ including  and excluding the effect of double-color-filtering, respectively. The dotted curve demonstrates rapidity dependence of the ISI effects at RHIC. It is calculated at $y=2$ and is to be compared with the upper solid curve at $y=0$.
}
 \end{figure}

Now we are in a position to perform realistic calculations for the nuclear suppression factor in $AB$ collisions. Provided that the $\bar cc$ production occurs in the long coherence length regime for both nuclei, the nuclear suppression factor at impact parameter $b$ reads,

\beqn
R_{AB}(b)&=&{1\over T_{AB}(b)}\int d^2\tau\,\frac{T_A(\tau)T_B((\vec b-\vec\tau)}
{(\Lambda_A^+-\Lambda_A^-)(\Lambda_B^+-\Lambda_B^-)}\,
\nonumber\\&\times&
\ln\left[\frac{(1+\Lambda_A^-+\Lambda_B^+)(1+\Lambda_A^++\Lambda_B^-)}
{(1+\Lambda_A^++\Lambda_B^+)(1+\Lambda_A^-+\Lambda_B^-)}\right]
\label{280}
\eeqn
where
\beqn
\Lambda_A^+&=&{\la r_T^2\ra\over2}\,C(E^A_{\bar cc})T_A(\tau);
\label{300a}\\
\Lambda_A^-&=&{7\la r_T^2\ra\over32}\,C(E^A_{\bar cc})T_A(\tau);
\label{300b}\\
\Lambda_B^+&=&{\la r_T^2\ra\over2}\,C(E^B_{\bar cc})T_B(\vec b-\vec\tau);
\label{300c}\\
\Lambda_B^-&=&{7\la r_T^2\ra\over32}\,C(E^B_{\bar cc})T_B(\vec b-\vec\tau);
\label{300d}
\eeqn
and $E^{A,B}_{\bar cc}$ are the energies of the $\bar cc$ in the rest frames of the nuclei $A$ and $B$ respectively. The result of Eq.~(\ref{280}) is plotted as the upper solid curve in Fig.~\ref{A-A-b}. 
For comparison, the result of conventional calculations assuming simple multiplication of the suppression factors in the two nuclei, is depicted by the dashed curve. We see that the mutual color filtering makes the nuclei considerably more transparent. This effect should be more prominent for production of $\Psi'$ and $\chi$. 

With Eq.~(\ref{280}) we can trace the $y$-dependence of $R_{AA}$. It turns out to be rather weak
at the energy of RHIC, due to the approximate linearity of $y$-dependence in $pA$ depicted in Fig.~\ref{d-A-y}, which leads to a compensation of the nuclear effects in the colliding nuclei. However, at sufficiently large $y$, say $y=2$,
the condition of long coherence length breaks down in one of the nuclei. Then the $\bar cc$ dipole size is not frozen by Lorentz time delation, and the filtering in this particular nucleus is not effective any more. In this case the conventional multiplicative procedure is applicable, but the suppression factor in one nucleus (high $E_{\bar cc}$) should be calculated  differently, for a short $t_c$ regime. The result of such calculation is plotted by the bottom solid curve in Fig.~\ref{d-A-b}. We see that the nuclear suppression at $y=2$ is stronger than at $y=0$. This happens due to disappearance of the double-color-filtering effect.

\subsection{Boosting the saturation scale in $AA$ collisions}

Another mechanism which violates the conventional multiplicative procedure for the transition from $pA$ to $AA$ collisions, is the mutual boosting of the saturation scale in the colliding nuclei . It is controlled by the following reciprocity equations \cite{boosting},
\beq
\tilde Q_{sB}^2(x_B)=\frac{3\pi^2}{2}\alpha_s(\tilde Q_{sA}^2+Q_0^2)
x_B g_N(x_B,\tilde Q_{sA}^2+Q_0^2)\,T_B;
\label{320a}
\eeq
 \beq
\tilde Q_{sA}^2(x_A)=\frac{3\pi^2}{2}\alpha_s(\tilde Q_{sB}^2+Q_0^2)
x_A g_N(x_A,\tilde Q_{sB}^2+Q_0^2)\,T_A,
\label{320b}
\eeq
where we consider a collision of two raws of nucleons $T_A$ and $T_B$, and production 
of a heavy quark pair with fractional momenta $x_A$ and $x_B$ relative to the colliding nucleons.
In what follows we consider heavy quarkonium production at forward rapidities relative to the momentum direction of nucleus $A$, which we call the beam. Correspondingly, the target nucleus is $B$. The values of $x_{A(B)}$ in (\ref{320a}) and (\ref{320b}) for charmonium production with positive rapidity $y$
are calculated as,
\beq
x_{A(B)}=\frac{\sqrt{M_{\J}^2+\la p_T^2\ra}}{\sqrt{s}}\,e^{\pm y}.
\label{330}
\eeq
The mean transverse momentum squared of $\J$ produced in $pp$ collisions at the energies of RHIC and LHC are $\la p_T^2\ra\approx 4\GeV^2$ and $7\GeV^2$ respectively.

The gluon distribution function $g_N(x,Q^2)$ contains the parameter $Q_0$ needed to regularize the infra-red behavior adjusting the saturation momentum in $pA$ collision to the known value \cite{jkt},
\beqn
Q_{sA}^2(b,E_{\bar QQ})&=&
\vec\nabla_{r_T}^2\,\sigma_{dip}(r_T,E_{\bar QQ})\Bigr|_{r_T=0}\,T_A(b)
\nonumber\\ &=&
2C(E_{\bar QQ})\,T_A(b),
\label{340} 
\eeqn
where parameter $C(E_{\bar QQ})$ introduced in Eq.~(\ref{140}) is well fixed by HERA data.
Solution of Eq.~(\ref{340}) leads to the infra-red cutoff parameter $Q_0^2\approx 1.7\GeV^2$, which is nearly independent of energy.

Solution of equations (\ref{320a})-(\ref{320b}) shows that the modified saturation scales $\tilde Q_{sA(B)}^2$ in $AB$ collisions considerably exceed the conventional scales Eq.~(\ref{340}) relevan for $pA$ collisions. The essential point of this consideration that the boosted saturation scale leads to an increase of the break-up cross section of a dipole, because the saturation momentum is directly related to the factor $C(E)$ in the dipole cross section, Eq.~(\ref{340}). Thus, the break-up cross section $\sigma_{dip}(r_T)$ for a dipole propagating through the nucleus $B$
modifies as
\beq
\sigma_{dip}(r_T)
\,\Rightarrow \, \tilde\sigma^B_{dip}(r_T)=K_{A}\,\sigma_{dip}(r_T),
\label{360}
\eeq
where
\beq
K_A=\frac{\tilde Q_{sA}^2}{Q_{sA}^2}.
\label{370}
\eeq
This boosting factor for the dipole absorption cross section in the nucleus $B$ is controlled by the boosted saturation scale of the nucleus $A$, which
implicitly depends on $T_A$, $T_B$, $x_A$, and $x_B$ in accordance with the equations (\ref{320a})-(\ref{320b}). Correspondingly the boosted break-up cross section of the $\bar cc$ dipole
propagating through the nucleus $A$ is given by $\tilde\sigma^A_{dip}(r_T)=K_B\,\sigma_{dip}(r_T)$.

\begin{figure}[htb]
\begin{center}
\includegraphics[width=6.5cm]{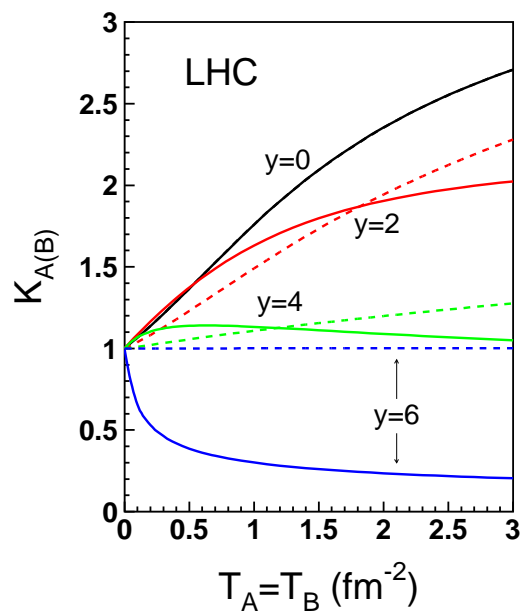}
\end{center}
\caption{\label{KA-KB-lhc} (Color online) The boosting factors $K_A(x_A)$ (solid curves) and $K_B(x_B)$ (dashed curves) defined in (\ref{370})
and calculated with the reciprocity equations (\ref{320a})-(\ref{320b}) for $\sqrt{s}=5.5\TeV$ with $x_{A(B)}$ defined in (\ref{330}). Each pair of curves is marked by the rapidity for which it is calculated. 
 }
 \end{figure}

We calculated the factors $K_A$ and $K_B$ solving the reciprocity equations (\ref{320a})-(\ref{320b})
supplied with the MSTW2008 code \cite{mstw} for the gluon distributions. The results at $\sqrt{s}=5.5\TeV$ are plotted in Fig.~\ref{KA-KB-lhc} as function of $T_A=T_B$ (central $AA$ collision) for different rapidities of the produced charmonium. 
Solid and dashed curves present $K_A$ and $K_B$ respectively 

We see that at the boosting factors are maximal at $y=0$, where $K_A\equiv K_B$. At forward rapidities $x_A$ rises, while $x_B$ decreases. As a result, the boosting factors $K_A$ and $K_B$ remain similar, and both are falling with rapidity. Eventually, at large rapidity $y=6$ the projectile $x_A$ becomes so large that the shift of the scale does not lead to an increase of the gluons density in
$A$, but makes it smaller \cite{saturation}. Therefore the effect of boosting turns into a suppression, i.e. the nucleus $A$ in this case becomes more transparent, rather than opaque, for $\J$.

\begin{figure}[htb]
\begin{center}
\includegraphics[width=6.5cm]{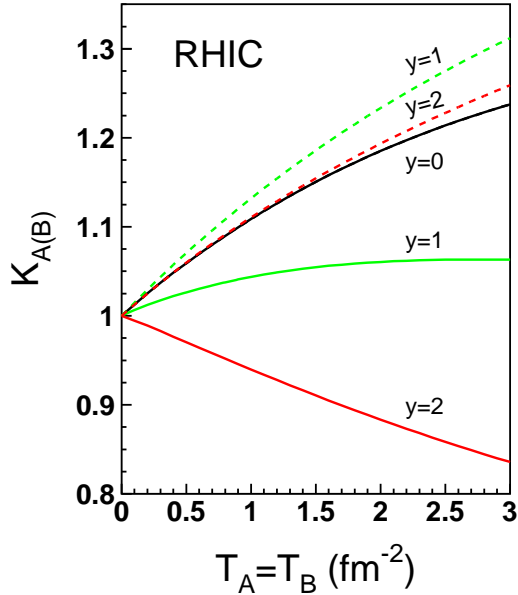}
\end{center}
\caption{\label{KA-KB-rhic} (Color online) The same as in Fig.~\ref{KA-KB-lhc}, but at $\sqrt{s}=200\GeV$. 
 }
 \end{figure}

Similar calculations at $\sqrt{s}=200\GeV$ depicted in Fig.~\ref{KA-KB-rhic} demonstrate a different behavior.
Surprisingly, while $K_B$ is falling with rapidity, as expected, $K_A$ is rising. This rise of the boosting factor is related to steep drop at forward rapidities of the saturation momentum in the projectile nucleus $A$, the denominator of (\ref{370}). In fact, the very existence of a saturation regime at such large $x_Q$ is questionable.

We see that the boosting effect significantly increases the saturation scales in colliding heavy nuclei,
especially at the mid rapiditiy and high energies. The rapidity dependence of the boosting factor is more complicated, and is related to different variation with scale of the gluon distribution function at different values of Bjorken $x$.
It worth reminding that what is plotted in Figs.~\ref{KA-KB-lhc} - \ref{KA-KB-rhic} are the ratios. The absolute value of the saturation momentum  in the nucleus $B$ ($A$) is steeply rising (falling) with rapidity.

As far as the transparency of nuclear medium  for heavy quark dipoles in the case of nuclear collisions is different from one measured in $pA$ collision  the simplified multiplicative prescription of \cite{phenix-dA,cassagnac,kps-psi} for $pA$ to $AA$ transition may be quite incorrect.

\begin{figure}[htb]
\begin{center}
\includegraphics[width=6.5cm]{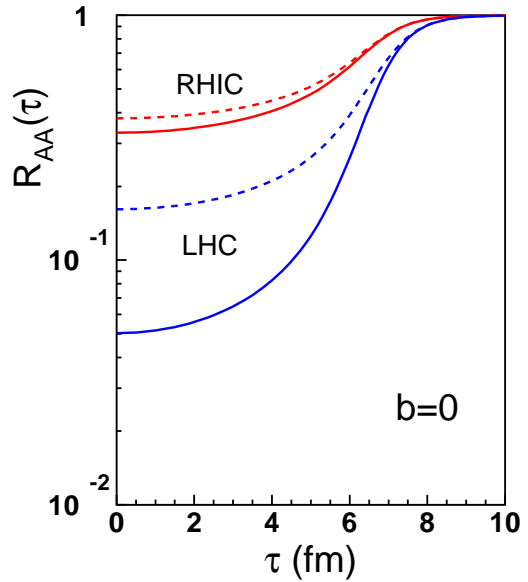}
\end{center}
\caption{\label{A-A-tau} (Color online) Effect of the boosted saturation scale on the nuclear ratio for $\J$ production in central ($b=0$) $Au$-$Au$ collisions at $y=0$ as function of impact parameter $\tau$. The upper and bottom dashed curves correspond to $\sqrt{s}=200\GeV$ and $5.5\TeV$ respectively. They are calculated in the same way as the solid curves in Fig.~\ref{A-A-b}. The solid curves here are calculated with the boosted saturation scale, which makes the nuclei more opaque for heavy dipoles.
}
 \end{figure}

In Fig.~\ref{A-A-tau} we demonstrate  the strength of the boosting effect on the production rate of $\J$ in central ($b=0$) $Au$-$Au$ collision as function of impact parameter $\tau$. 
The dashed curves include the double-color-filtering effect, but exclude the saturation scale boosting, which is added to produce the solid curves. The solid curves include the effect of boosted saturation scale.
The upper and bottom pairs of curves correspond to the energies of RHIC and LHC respectively. 
We conclude that $\J$ should be  significantly stronger suppressed in $AA$ collisions, than usually expected extrapolating from $pA$, and one should not misinterpret this suppression as anomalous effect related to FSI with the dense medium. Notice that a stronger suppression of $\J$ in the ISI stage of the collision compared to the one used in \cite{kps-psi}   
should result in even smaller transport coefficient of the dense medium extracted  from RHIC data on $\J$ production. 

\subsection{Boosted gluon shadowing}

As we have already mentioned, in terms of the Fock decomposition gluon shadowing corresponds to multiple interactions of higher Fock states, containing gluons. One may wonder if the double color filtering effect can affect the amount of gluon shadowing in nuclear collisions. The answer is not.
The lifetime of such a fluctuation produced by a nucleon in the nucleus $A$ may be sufficiently long only relative to the nucleus $B$, but is very short relative to the parent nucleus $A$. Therefore no gluonic fluctuations undergo double color filtering. In terms of Gribov inelastic shadowing this means that the diffractive excitation of the nucleons of $A$ propagate through $B$ independently of
the excitations of $B$ propagating through $A$.

Thus, the gluon shadowing factors factorize in $AB$ collisions,
\beq
R_g^{AB}(\vec b,\vec\tau)=R_g^A(\tau)\,R_g^B(\vec b-\vec\tau).
\label{380}
\eeq

Even if gluon shadowing for heavy quarkonium production in $pA$ collision is known, say, from analyses of DIS data \cite{DS}, it is integrated over impact parameter, while  in (\ref{380}) one needs to know its impact parameter dependence in order to predict gluon shadowing in heavy ion collisions. Already this
problem is a serious obstacle for extrapolation from $pA$ to $AA$.

Aside from the ad hoc parametrizations of the $b$-dependence existing in the literature, one has to rely on a fully developed theoretical model for gluon shadowing to predict its $b$-dependence.
Unfortunately, no satisfactory theoretical description, which would  work at all kinematic regimes, has been developed so far. The most rigorous quantum-mechanical treatment of gluon shadowing within the path-integral technique \cite{kst2,krtj} is the lowest order calculation, which might be a reasonable approximation only for light nuclei, or for the onset of shadowing. Contribution of higher Fock components is still a challenge. This problem has been solved only in the limit of long coherence lengths for all radiated gluons, in the form known as the Balitsky-Kovchegov equation (BK) \cite{b,k}.
Numerical solution of this equation is quite complicated and includes lot of modeling \cite{wiedemann}.
A simpler equation, which only employs a modeled shape of the saturated gluon distribution, was derived in \cite{saturation}. It leads to a gluon distribution in nuclei, which satisfies the unitarity bound \cite{bound}, and is quite similar to the numerical solutions of the BK equation. The equation reads \cite{saturation},
\beq
R_g=1-
\frac{R_g^2\,n_0^2\,n_{eff}}{(1+R_g\,n_0)^2(1+n_{eff})}
\label{400}
\eeq
where 
\beqn
n_0(E_{\bar cc},b)&=& \frac{9\,C(E_{\bar cc})}{2\,Q_{qN}^2(E_{\bar cc})}\,T_A(b);
\nonumber\\
n_{eff}(E_{\bar cc},b)&=&{9\over4}\,C(E_{\bar cc})\,r_0^2\,T_A(b).
\label{420}
\eeqn
The energy dependent factor $C(E_{\bar cc})$ was introduced above in (\ref{140}). The mean size of a gluonic dipole, $r_0\approx 0.3\fm$, is dictated by data \cite{kst2,spots}. 
We rely on the saturated shape of the dipole-nucleon cross section with the saturation scale $Q_{qN}(E_{\bar cc}) = 0.19\GeV\times(E_{\bar cc}/1GeV)^{0.14}$, fitted to DIS data \cite{kst2,saturation}. One can switch from energy to Bjorken $x$ dependence in these equations,
using the relation Eq.~(\ref{330}) and replacing $s\Rightarrow 2E_{\bar cc}m_N$.

Notice that Eq.~(\ref{400} does not contain any hard scale, but only a semi-hard one controlled by $r_0$ \cite{spots}. Therefore its solution should be treated as the starting gluon distribution
at the semi-hard scale $Q^2\approx 4/r_0^2\approx Q_)^2$, where $Q_0$ was introduced above in 
(\ref{320a})-(\ref{320b}). Shadowing for heavy quarkonium production should be DGLAP evolved up to an appropriate hard scale.

The effect of boosted saturation scale leads to a modification of the factor $C(E_{\bar cc})$
which is different for nuclei $A$ and $B$,
\beq
C(E_{\bar cc})\,\Rightarrow\, \tilde C_{A(B)}(E_{\bar cc})=K_{A(B)}\,C(E_{\bar cc}).
\label{440}
\eeq
Solving equation (\ref{400}) with such a boosted saturation scale one arrives at a modified nuclear ratio for gluons in nuclei $A$ and $B$,
$\tilde R^{A(B)}_g(E_{\bar cc},b)$, for which the factorized relation (\ref{380}) can be used,
\beq
\tilde R_g^{AB}(\vec b,\vec\tau)=\tilde R_g^A(\tau)\,\tilde R_g^B(\vec b-\vec\tau).
\label{460}
\eeq

\begin{figure}[htb]
\begin{center}
\includegraphics[width=6.5cm]{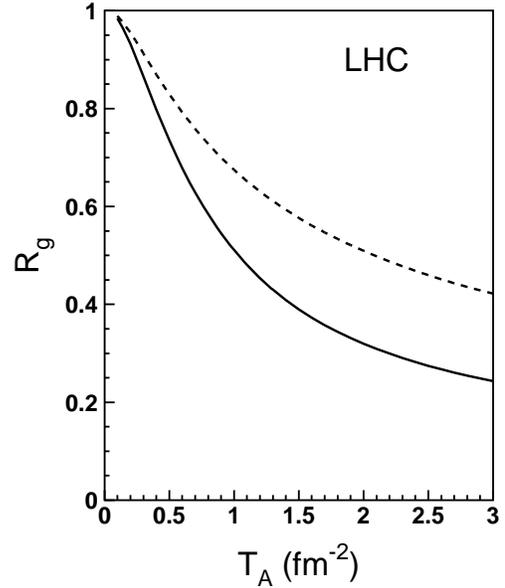}
\end{center}
\caption{\label{shad-lhc}  Gluon nuclear ratio $R_g$ as function of nuclear thickness $T_A$ calculated with equation (\ref{400}) at the semihard scale $Q_0$. The dashed curve presents gluon shadowing corresponding to
hadron-nucleus collisions. The solid curve include the boosting effects specific for central nucleus-nucleus collisions ($T_A=T_B$). Shadowing is calculated at the starting scale $Q_0$ and should be evolved up to a higher scale.}
 \end{figure}

A numerical example for the boosting effect on gluon shadowing is depicted in Fig.~\ref{shad-lhc}.
The dashed and solid curves show the solutions of Eq.~(\ref{400}) without and with inclusion of the boosting effects,
calculated for $\J$ produced with $y=0$ at $\sqrt{s}=5500\GeV$. We see that the boosting effect is considerable. As was mentioned above, shadowing is calculated at the starting scale $Q_0$ and should be evolved up to a proper higher scale.

\section{Summary and prospectives}

The dipole formalism based in the universal dipole cross section well fitted to HERA data, successfully  describes the propagation of heavy quark dipoles through cold nuclear matter. Our parameter free calculation presented in Fig.~\ref{d-A-y} well agree with data from RHIC for $\J$ production in $d$-$Au$ collisions. We do all calculations within the "frozen" approximation, assuming that the dipole size does not fluctuate during propagation through the nucleus. This condition is satisfied provided that the coherence time is long, $t_c\gg R_A$, what is well justified at the energies of RHIC and LHC.

We identify the main source of the observed nuclear suppression as a combined effect of break-up of the produced $\bar cc$ dipole in nuclear medium and charm quark shadowing. Although both are the high twist effects, their contribution to the observed nuclear suppression considerably exceeds
the effect of leading twist gluon shadowing. The latter is expected to be rather weak, basing on either theoretical predictions \cite{kst2}, or the NLO analysis \cite{DS} of DIS data. 

Even if the nuclear effects for heavy quarkonium production are understood, the transition to $AA$ collisions is rather complicated. Several new phenomena, specific for nuclear collisions, make such a transition model dependent. The first one, double color filtering, makes the nuclei more transparent for quark dipoles than in $pA$ collisions. The second one is the mutual boosting of the saturation scales in the colliding nuclei, which makes the nuclei more opaque for quark dipoles. Although these two effects act in opposite directions, the latter is much stronger, as one can see in Fig.~\ref{A-A-tau}.

Another direct consequence of the increased saturation scale in nuclear collisions is boosting of gluon shadowing, which violates the factorized relation, Eq.~(\ref{380}).  Gluon shadowing 

A direct way to access the saturation scale experimentally is to measure transverse momentum broadening of charmonia produced on nuclei.  Then the boosting effect should show up as an increase of broadening of $\J$ produced in $AA$ compared to $pA$ collisions. Indeed such an effect was clearly observed in the NA60 and NA50 experiments at $E_{lab}=158\GeV$ \cite{na60}. The broadening in nuclear collisions was found twice as large as in $pA$ measurements aith the same path length in the nuclear medium. Within the rather good statistical and systematic accuracy, the effect is quite certain.
The magnitude of the observed boosting is larger than is predicted by the equations (\ref{320a})-(\ref{320b}) at this energy, so this problem needs further study, and the data are to be confirmed by independent measurements.

Although we considered here only the high energy limit of "frozen" dipoles, the $pA$ to $AA$ transition is not trivial at medium high energies, say at SPS, either \cite{k-bnl}. 
This case will be also investigated and published elsewhere. 


\begin{acknowledgments}

We are thankful to Sasha Tarasov for numerous informative discussions. 
This work was supported in part
by Fondecyt (Chile) grants 1090291, 1090236, and 1100287, by DFG
(Germany) grant PI182/3-1, and by Conicyt-DFG grant No. 084-2009.

\end{acknowledgments}


\begin{thebibliography}{99}

\bibitem{satz}
  T.~Matsui and H.~Satz,
  Phys.\ Lett.\  B {\bf 178}, 416 (1986).

\bibitem{kps-psi}
  B.~Z.~Kopeliovich, I.~K.~Potashnikova and I.~Schmidt,
 Phys.\ Rev.\  C {\bf 82}, 024901 (2010);
  arXiv:1006.3042 [nucl-th].

\bibitem{k-bnl}
  B.~Z.~Kopeliovich,
  arXiv:1007.4513 [hep-ph].

\bibitem{gbw} K.~J.~Golec-Biernat and M.~Wusthoff,
  Phys.\ Rev.\  D {\bf 60}, 114023 (1999).

\bibitem{kth-psi}
  B.~Kopeliovich, A.~Tarasov, J.~H\"ufner,
  Nucl.\ Phys.\  A {\bf 696}, 669 (2001).
  
\bibitem{bl}
  S.~J.~Brodsky and J.~P.~Lansberg,
  Phys.\ Rev.\  D {\bf 81}, 051502 (2010).

\bibitem{klt}
  D.~Kharzeev, E.~Levin, M.~Nardi and K.~Tuchin,
  Nucl.\ Phys.\  A {\bf 826}, 230 (2009).
    
\bibitem{kz91}
  B.~Z.~Kopeliovich and B.~G.~Zakharov,
  Phys.\ Rev.\  D {\bf 44}, 3466 (1991).

 \bibitem{ws} H. de Vries, C.W. de Jager, C. de Vries : At. Data Nucl. Data Tabl. 36, 495 (1987).

 \bibitem{phenix-dA}
  A.~Adare {\it et al.}  [PHENIX Coll.],
  Phys.\ Rev.\  C {\bf 77}, 024912 (2008).

\bibitem{krt2}
  B.~Z.~Kopeliovich, J.~Raufeisen and A.~V.~Tarasov,
  Phys.\ Rev.\  C {\bf 62}, 035204 (2000).

\bibitem{kst2} B.Z.~Kopeliovich, A.~Sch\"afer and A.V.~Tarasov, Phys.
Rev. {\bf D62} (2000) 054022.

\bibitem{spots}
  B.~Z.~Kopeliovich, I.~K.~Potashnikova, B.~Povh and I.~Schmidt,
  Phys.\ Rev.\  D {\bf 76}, 094020 (2007).
   
\bibitem{brahms}
  B.~Z.~Kopeliovich, J.~Nemchik, I.~K.~Potashnikova, M.~B.~Johnson and I.~Schmidt,
  Phys.\ Rev.\  C {\bf 72}, 054606 (2005).
  
 \bibitem{DS} D.~de~Florian and R.~Sassot, Phys. Rev. D {\bf 69},
074028 (2004).

\bibitem{cassagnac}
  R.~Granier de Cassagnac,
  J.\ Phys.\ G {\bf 34}, S955 (2007).

\bibitem{zkl}
  B.~Z.~Kopeliovich, L.~I.~Lapidus and A.~B.~Zamolodchikov,
  JETP Lett.\  {\bf 33}, 595 (1981)
  [Pisma Zh.\ Eksp.\ Teor.\ Fiz.\  {\bf 33}, 612 (1981)].

\bibitem{bbgg}
  G.~Bertsch, S.~J.~Brodsky, A.~S.~Goldhaber and J.~F.~Gunion,
  Phys.\ Rev.\ Lett.\  {\bf 47}, 297 (1981).

\bibitem{boosting}
  B.~Z.~Kopeliovich, H.~J.~Pirner, I.~K.~Potashnikova and I.~Schmidt,
  arXiv:1007.1913 [hep-ph].

\bibitem{jkt}
  M.~B.~Johnson, B.~Z.~Kopeliovich and A.~V.~Tarasov,
  Phys.\ Rev.\  C {\bf 63}, 035203 (2001).
  
\bibitem{mstw}
  A.~D.~Martin, W.~J.~Stirling, R.~S.~Thorne and G.~Watt,
  Eur.\ Phys.\ J.\  C {\bf 63}, 189 (2009);
   Eur.\ Phys.\ J.\  C {\bf 64}, 653 (2009).

\bibitem{saturation}
  B.~Z.~Kopeliovich, I.~K.~Potashnikova and I.~Schmidt,
  Phys.\ Rev.\  C {\bf 81}, 035204 (2010).
  
\bibitem{krtj}
  B.~Z.~Kopeliovich, J.~Raufeisen, A.~V.~Tarasov and M.~B.~Johnson,
  Phys.\ Rev.\  C {\bf 67}, 014903 (2003).

\bibitem{b} I. Balitsky, Nucl. Phys. B 463, 99 (1996).

\bibitem{k} Y. V. Kovchegov, Phys. Rev. D 60, 034008 (1999).

\bibitem{wiedemann}
  J.~L.~Albacete, N.~Armesto, J.~G.~Milhano, C.~A.~Salgado and U.~A.~Wiedemann,
  Phys.\ Rev.\  D {\bf 71}, 014003 (2005).

 \bibitem{bound}
  B.~Z.~Kopeliovich, E.~Levin, I.~K.~Potashnikova and I.~Schmidt,
  Phys.\ Rev.\  C {\bf 79}, 064906 (2009).

 \bibitem{na60}
  E.~Scomparin  [NA60 Coll.],
  Nucl.\ Phys.\  A {\bf 830}, 239C (2009);
    R.~Arnaldi  [NA60 Coll.],
  Nucl.\ Phys.\  A {\bf 830}, 345C (2009).

\end{thebibliography}
\end{document}